\documentclass[aps,prb,twocolumn,groupedaddress,showpacs,10pt]{revtex4-1}
\usepackage{graphicx}
\usepackage{color}
\usepackage{calc}
\usepackage{array}
\usepackage{amsmath, amssymb}
\usepackage{natbib}
\usepackage{hyperref}

\newcommand{\red}[1]{{\color{black}{#1}}}

\begin{document}

\title{Topological superconductivity from first-principles I: \\
Shiba band structure and topological edge states of artificial spin chains}


\author{Bendeg\'uz Ny\'ari$^{1,2}$} 
\author{Andr\'as L\'aszl\'offy$^{3}$}
\author{G\'abor Csire$^{4,5}$}
\author{L\'aszl\'o Szunyogh$^{1,2}$}
\author{Bal\'azs \'Ujfalussy$^{3}$}

\affiliation{$^1$Department of Theoretical Physics, Institute of Physics, Budapest University of Technology and Economics, M\H uegyetem rkp.~3., HU-1111 Budapest, Hungary}
\affiliation{$^2$ELKH-BME Condensed Matter Research Group, Budapest University of Technology and Economics, M\H uegyetem rkp.~3., HU-1111 Budapest, Hungary}
\affiliation{$^3$Wigner Research Centre for Physics, Institute for Solid State Physics and Optics, H-1525 Budapest, Hungary}
\affiliation{$^4$Materials Center Leoben Forschung GmbH, Roseggerstraße 12, 8700 Leoben, Austria.}
\affiliation{$^5$Catalan Institute of Nanoscience and Nanotechnology (ICN2), CSIC, BIST, Campus UAB, Bellaterra, Barcelona, 08193, Spain}

\date{\today}
\begin{abstract}
Magnetic chains on superconductors hosting Majorna Zero Modes (MZMs) attracted high interest due to their possible applications in fault-tolerant quantum computing. 
However, this is hindered by the lack of a detailed, quantitative understanding of these systems.
As a significant step forward, we present a first-principles computational approach based on a microscopic relativistic theory of inhomogeneous superconductors applied to an iron chain on the top of Au-covered Nb(110) to study the Shiba band structure and the topological nature of the edge states.
Contrary to contemporary considerations, our method enables the introduction of quantities indicating band inversion without fitting parameters in realistic experimental settings, 
\red{
holding thus the power to determine the topological nature of zero energy edge states in an accurate ab-initio based description of the experimental systems.} 
We confirm that ferromagnetic Fe chains on Au/Nb(110) surface do not support any separated MZM; however, a broad range of spin-spirals can be identified with robust zero energy edge states displaying signatures of MZMs. For these spirals, we explore the structure of the superconducting order parameter shedding light on the internally antisymmetric triplet pairing hosted by MZMs.  We also reveal a two-fold effect of spin-orbit coupling: although it tends to enlarge the topological phase regarding spin spiraling angles, however, it also extends the localization of MZMs.
Due to the presented predictive power, our work fills a big gap between the experimental efforts and theoretical models while paving the way for engineering platforms for topological quantum computation.
\end{abstract}

\maketitle

\section{Introduction}
\red{Topological superconductivity is an exotic state of matter where the condensate of Cooper pairs of electrons spontaneously breaks the U(1) gauge symmetry and simultaneously 
exhibit a non-trivial topological gap structure\cite{Liang2011, Sato2017}.}
Although there may well be materials that develop intrinsic topological superconductivity providing natural platforms for MZMs\cite{Wray2010, Sasaki2011, Ueno2013, Zhang2018, Ran2019, Jin2019, Avers2020, Jiao2020, Li2021, Badger2022}, the real breakthrough $-$ that has created a great number of routes to such platforms $-$ was the realization that one can create topological superconductivity based on artificial heterostructures\cite{Fu2008, Lutchyn2010, oreg2010helical, choy2011majorana, Mourik2012, Kjaergaard2012, Martin2012, NadjPerge2013, klinovaja2013topological, Rainis2013, Vazifeh2013, nakosai2013two, klinovaja2013topological, li2014topological, heimes2014majorana, NadjPerge2014, Ruby2015, hui2015majorana, dumitrescu2015majorana, sarma2015substrate, Peng2015, christensen2016spiral, fatin2016wireless, li2016manipulating, Deng2016, Hell2017, marra2017controlling, Mashkoori2020, Cayao2020, Cayao2021, Flensberg2021, Beck2021, Marra2022, Marra2022review, gungordu2022majorana, steffensen2022topological, NeuhausSteinmetz2022, crawford2023increased}.
Due to the bulk-edge correspondence principle\cite{Liang2011}, topological superconductivity is manifested in zero energy edge states: the renowned Majorana Zero Modes (MZMs).
These states have drawn a significant interest of the scientific community since MZMs bear the potential application for topological quantum computing\cite{Nayak2008non,Alicea2012, Beenakker2020}.
However, MZMs in superconducting heterostructures are still elusive because it is very difficult to uniquely identify them experimentally. Several promising STM experiments have been performed on various systems, which show peaks in the differential conductivity at zero energy\cite{Jeon2017,Kim2018,Schneider2021a,Schneider2021b} in the superconducting gap of the host. However, this does not impose a strict evidence that the observed states at the end of the chain are indeed the long sought MZMs and further information about the nature of these peaks are difficult to obtain. 

To address this problem we developed a first-principles based computational approach serving as a template for detailed analysis of different MZM platforms. This is allowed by the Green's function based solution of the Kohn--Sham--Dirac--Bogoliubov--de Gennes (KSDBdG) equations\cite{csire2018relativistic,Nyari2021}.
On one hand, it has been demonstrated previously, that many aspects of the STM experiments are reproducible by such calculations\cite{Nyari2021, saunderson2022full, park2023effects, wu2023magnetic}, while on the other hand, it allows calculating other quantities, like spin-polarization and the superconducting order parameter (OP)\cite{Linscheid2015}, which are important to understand the nature of these states. Furthermore, some of their properties can be further explored and tested by computational experiments which go beyond the capabilities of conventional experimental techniques. 

In what follows, we show first principles calculations performed for Fe chains on top of superconducting Nb(110) host with a single epitaxial Au overlayer, as introduced in Ref.~\onlinecite{beck2023search}, in the superconducting state, where \red{relativistic effects}, superconductivity and the complex electronic structure is treated on the same level. Based on previous simple model calculations there are two essential ingredients for the formation of MZMs in spin chains proximitized with s-wave superconductors, a strong Rashba spin-orbit coupling (SOC)\cite{Lutchyn2010,oreg2010helical} or a non-collinear spin structure, like a spin spiral\cite{NadjPerge2013, Pientka2013, Pientka2015, heimes2015interplay, Marra2022review}, both inducing $p$-wave pairing\cite{Alicea2011} and hence topological superconductivity. One idea, which was realized in a recent experiment is to cover the surface of an s-wave superconductor with a single atomic layer of a heavy-metal\cite{beck2023search}. This has the advantage of keeping the relatively large superconducting gap of Nb, while simultaneously enhancing the SOC in the system. However, in spite of the enhanced SOC there was no experimentally observable minigap in the system. Further theoretical investigations revealed\cite{beck2023search}, that, by forcing the system into a 90$^\circ$ spin-spiral state, it is possible to open up a minigap hosting zero energy end states. 
This previously obtained finding just asks for the application of first principles methods described in Section~\ref{sec:method},
which \red{can further substantiate the topological classification of these states} and provide practical guidance for further experiments. 
In Section~\ref{sec:spiral} by considering a wide range of spiraling angles we make quantitative predictions for the local density of states (LDOS) as described in Ref.~\onlinecite{Nyari2021} and by changing the spiraling angle we show it drives the system through topological phase transitions. At these points, the minigap closes and the zero energy states appear or disappear. We also show how this picture changes if we utilize the capability of our method to artificially scale the spin-orbit coupling (SOC).
\red{In an attempt to validate the developed method, we verify the expected result that in a ferromagnetic chain without spin-orbit coupling there is no topological superconductivity and MZMs\cite{NadjPerge2013}.} 
In Section~\ref{sec:localize} we shall study the spatial distribution of the zero energy peak which reveals a two-fold effect of SOC: although it tends to enlarge the topological phase regarding spin spiraling angles, however, it also extends the localization of MZMs.
Our model makes it possible to explore quantities that are beyond the current capabilities of experiments to measure. The superconducting order parameter (OP)\cite{Linscheid2015} belongs to this category and in Section~\ref{sec:op} we discuss that it has a more complex structure (involving both spin singlet and triplet parts) than in the well-known prototype models\cite{NadjPerge2013, li2014topological, Pyhnen2014, heimes2015interplay, Pientka2015, brydon2015topological, sarma2015substrate, christensen2016spiral, Flensberg2021} for Majorana zero modes. We identify that the structure of the superconducting OP (more precisely, its energy resolution, introduced later) can serve as an indicator of band inversion and thus topological superconductivity.
Finally in Section~\ref{sec:topology} we further analyze the topological nature of the minigap by illustrating the appearance of band inversion based on the spin singlet OP and the quasi-particle charge density of states.

\section{First principles based treatment of artificial spin chain on superconducting host}
\label{sec:method}

The density functional theory yielding Kohn--Sham equations is proven to successfully describe material-specific properties. The concept of superconductivity can be introduced into this theory by treating the superconducting OP as an additional (so-called) anomalous density\cite{oliveira1988density}. Such generalization of Kohn--Sham equations leads to the following KSDBdG Hamiltonian written in Rydberg units
\begin{equation}
H_{\text{DBdG}}=
 \begin{pmatrix}
   H_D & \Delta_{\text{eff}} \\
   \Delta_{\text{eff}}^\dagger & -H_D^*
 \end{pmatrix},
\end{equation}
where $ H_D(\vec r)=c\vec{\boldsymbol{\alpha}} \vec{p} + \left( \boldsymbol{\beta}-\mathbb{I}_4 \right) c^2/2+ \left( V_\text{eff}(\vec r)-E_F \right) \mathbb{I}_4 + \vec{\boldsymbol{\Sigma}}\vec{B}_\text{eff}(\vec r)$, with
$\vec{\boldsymbol \alpha} = \boldsymbol \sigma_x \otimes \vec{\boldsymbol \sigma}$, 
$\boldsymbol \beta = \boldsymbol \sigma_z \otimes \mathbb I_2$, 
$\vec{\boldsymbol \Sigma} = \mathbb I_2 \otimes \vec{\boldsymbol \sigma}$, 
$\vec{\boldsymbol \sigma} $ denotes the Pauli-matrices, 
and $\mathbb I_n $ being the identity matrix of order $n$. $V_{\text{eff}}(\vec r)$ and $\vec{B}_\text{eff}(\vec r)$  are the effective potential and the exchange field, respectively. $\Delta_\text{eff}(\vec r)$ is the effective $4 \times 4$ pairing potential matrix due to the four component Dirac spinors. The KSDBdG equations shall be solved self-consistently by assuming that the superconducting host has isotropic $s$-wave spin-singlet pairing as described by BCS theory\cite{bardeen1957theory}.

The central quantity of our approach, the Green's function, is obtained from the generalized multiple scattering theory (see Supplementary Note~1 for the detailed description of the method) in a self-consistent way. The great advantage of such a Green's function technique\cite{Minr2018} is the exact treatment of semi-infinite geometries (hence the superconducting host) together with the embedding of magnetic chains (see again Supplementary Note~1). \red{In this way, involving both the orbital and spin
degrees of freedom we can properly account for the microscopic complexity in the superconducting state of the studied iron nanowire placed on 
an Au monolayer grown epitaxially on the (110) surface of Niobium.}

For each site of the chain, the method yields the local Green's function matrix $\{ G^{nn,ab}_{Ls,L's'}(\varepsilon)\}$ 
(see Supplementary Note~1) where $n$ denotes the sites of the chain; $L=(l,m)$ and $L'=(l',m')$ are composed angular momentum indices; $s, s'$ are the spin indices; and $a, b$ corresponds to either the electron-like or the hole-like part of the Green's function. This quantity contains all information about the superconducting ground state involving the description of all the pairing states present in the system. Hence, this allows the calculation of the LDOS, and the energy-resolved OP related to different pairing states as defined later. Such an approach has two major advantages compared to effective models, like the tight-binding approximation. 
First, there is no further need to fit the electronic structure with artificial tight-binding parameters, which in turn allows for computational experiments with spin chains more easily. Second, it is crucial to have a proper model of the (semi-infinite) superconducting host if one aims to predict quantitatively the localization length of MZMs.
\red{The problem of insufficient modeling the semi-infinite host appears in most tight-binding approximations. 
These calculations resulted in an unrealistic gap to match the localization of MZMs\cite{Schneider2020, Crawford2022}, since the proximity-induced superconducting pairing was introduced into the chain as a parameter and not via an interaction with a superconducting host. 
The localization length is one of the most important quantities that decides whether the MZMs are separated enough to be feasible for topological quantum computation.
In the above context we mention that the host-induced suppression of Majorana localization length 
was studied on the model level by Das Sarma \textit{et al.}\cite{sarma2015substrate}
which also underlines the importance of the correct treatment of the host presented in this paper.}

\section{Spin-spirals in the superconducting state}
\label{sec:spiral}

\begin{figure}[!h]
 	\centering
     \includegraphics[width=\textwidth]{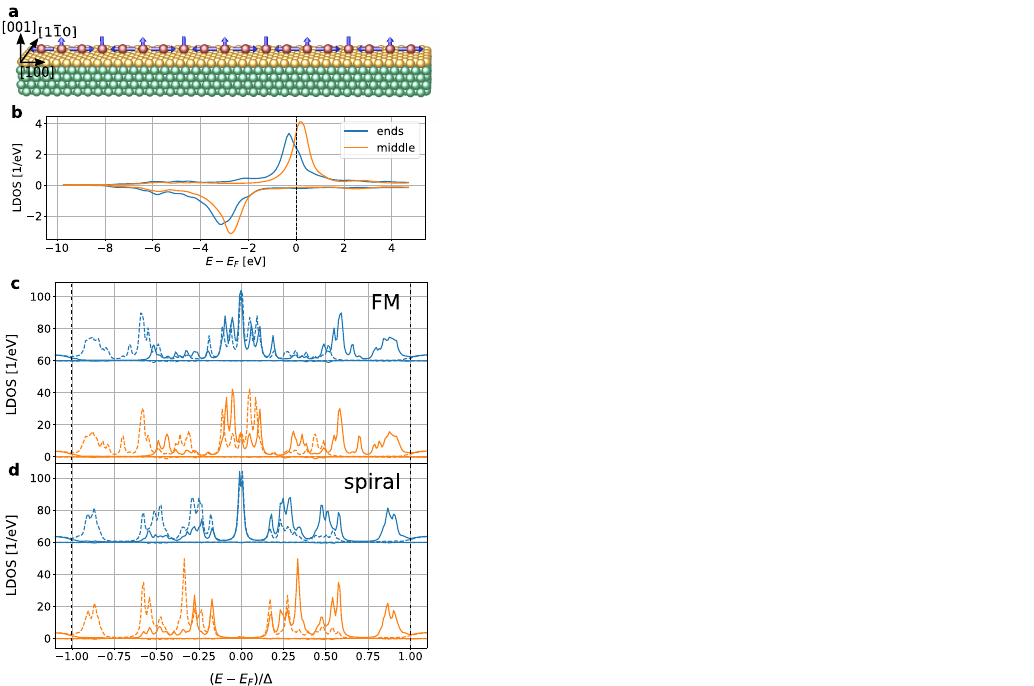}
     \caption{\label{fig_dos}The LDOS of the 2a-[100] Fe$_{19}$ chain on Au/Nb(110) in the normal and in the superconducting state. (a), the illustration of the 19 atomic 2a-[100] Fe chain on Nb(110) covered with a single monolayer Au, the spin configuration shows a Néel spiral with 90$^\circ$ spiraling angle. (b) the normal state local density of state for the ferromagnetic chain. (c) the LDOS in the superconducting state of the ferromagnetic chain. (d) the LDOS of the same chain as in (c) but in a Néel spiral state with 90$^\circ$ rotation angle as shown in panel (a). In panels (c) and (d) the solid lines are electron densities, while the dashed lines are hole densities and the blue curves are shifted with 60~1/eV. In all plots the positive values are from the minority spin channel and the negative values are from the majority spin channel. The blue curves are calculated on the first atom of the chain and the orange are from the middle of the chain. The black dashed vertical lines in panel (c) and (d) indicate the superconducting gap of the Nb, $\Delta=1.51$~meV.}
 \end{figure}
 
First, we discuss the results of the first principles calculations in the normal state on the same system introduced in Ref.~\onlinecite{beck2023search}, namely a 19-atom-long Fe chain with 2a (a=330 pm) nearest neighbor distance in the [100] direction as illustrated in Fig.~\ref{fig_dos}a (in short: 2a-[110] Fe$_{19}$ chain), placed on the (110) surface of an epitaxial Au monolayer covering the surface of Nb(110) described in details in Appendix A. The normal state LDOS is presented in Fig.~\ref{fig_dos}b for the ferromagnetic chain, with an out-of-plane magnetization. It can be seen that the majority spin channel is almost entirely filled and gives a negligible contribution to the normal state LDOS at the Fermi level, and an overwhelming contribution come from the minority spin channel. This fact is also expressed in the enhanced magnetic moment of about $3.8$ $\mu_B$, such elevated magnetic moments are typical for surface magnetic impurities. If we rather consider for example a $90^\circ$ Néel type spin spiral, as illustrated in Fig.~\ref{fig_dos}a, the normal state LDOS remains mostly unchanged, as discussed in Supplementary Note 3. In summary, we can not detect any feature of the normal state DOS that could signal the rather different behavior in the superconducting state we find later between the ferromagnetic and the spin-spiral state. 

Turning our attention to the superconducting state, first, we confirm the experimental finding\cite{beck2023search}, that the size of the induced gap in an Au overlayer system with a single atomic layer of Au on the Nb(110) surface does not differ from the size of the gap in pure Nb surface (the details are presented in Supplementary Note 4). 
Hereafter, we consider the 2a-[100] Fe$_{19}$ chain placed on this host system. The LDOS on some Fe impurity atoms of such a chain is plotted in Fig.~\ref{fig_dos}c for a ferromagnetic spin configuration, where the LDOS in the superconducting state is obtained by
\begin{equation}
    \mathrm{LDOS} (\varepsilon, n) = - \frac{1}{\pi} \Im~ \mathrm{Tr}~
                          \{ G^{nn,ab}_{Ls,L's'}(\varepsilon)\} \, .
\end{equation}
It can be seen that in the magnetic chain the Yu--Shiba--Rusinov states of the single Fe impurity hybridize within the superconducting gap of the host, as it was seen in the experiments\cite{Schneider2021a,Schneider2021b}, and the hybridized states occupy almost the entire energy range of the gap, including the zero energy. Although spin-orbit coupling naturally causes spin-mixing, it should be noted, that all states are, yet again, from one spin channel only, even though our calculations are fully relativistic.  This is not entirely surprising based on the normal state DOS of the chains (see Fig.~\ref{fig_dos}b) discussed previously. Most interestingly however, when we are repeating the calculation for a  90$^{\circ}$ Néel type spin spiral, the LDOS  plotted in Fig.~\ref{fig_dos}d, shows the opening of an internal gap of $\Delta_{\text{int}}=0.22$ meV around zero energy within the hybridized YSR states. Moreover, one peak appears right in the middle of this minigap, exactly at zero energy -- that is, at the Fermi energy -- on the atoms at both ends of the chain \red{which shall be referred as zero energy peak (\red{\red{ZEP}}).
In the context of scanning tunneling spectroscopy\cite{Choi2019} these peaks are manifested in the Zero Bias Peaks (ZBPs) observable in the differential conductance.}  Because these are exactly the features that are expected for a system with MZMs, it motivates to investigate other spin-spiral states via a computer experiment and look at how the MZMs and the minigap emerge as spin spiraling and SOC changes. But it has to be emphasized, that the fact that there are states at zero energy at some spiraling angle, does not necessarily mean that MZMs are found. As it has been demonstrated before, even in the case of a single magnetic impurity on the surface of a superconducting host, it is possible to obtain a state at zero energy by imposing a canting angle\cite{park2023effects}, even though it is not possible to obtain MZMs for a single impurity. Such states are just YSR states that are accidentally shifted to zero energy as a result of the canting angle. Therefore extreme caution and further analysis is needed regarding the classification of the minigap and the \red{\red{ZEP}}s and in the second paper in this series\cite{PRB-II} we will show how easy it is to obtain a ``fake'', or Quasi Majorana state.

\begin{figure*}[htb]
    \includegraphics[width=1\linewidth]{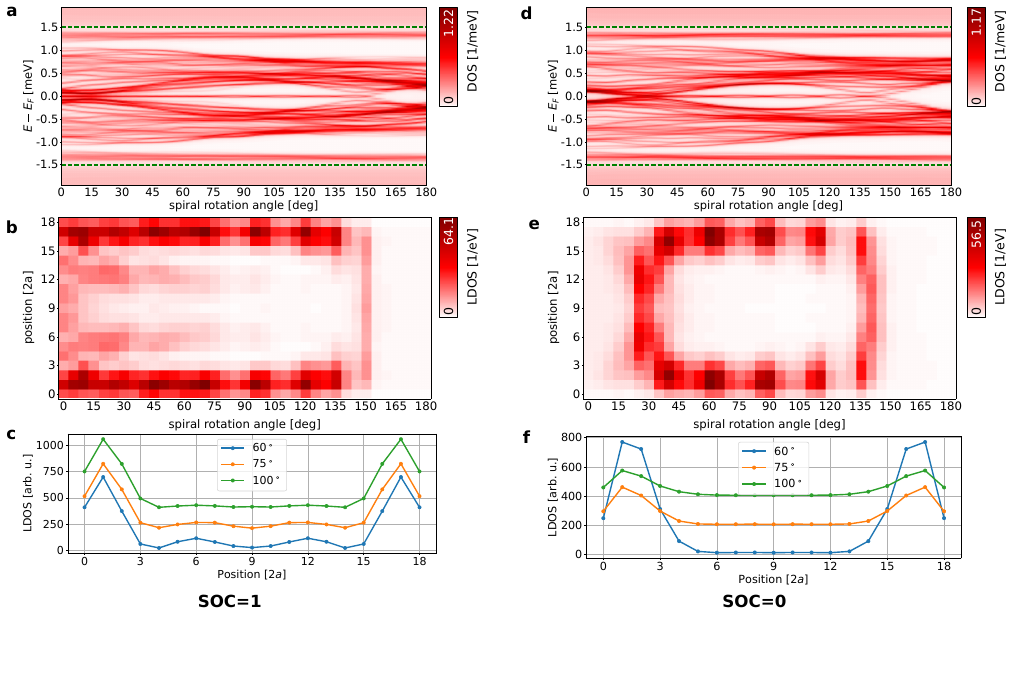}
    \vspace{-1.5cm}
     \caption{\label{fig_spiral} The effect of the SOC on the DOS, and the localization of the \red{\red{ZEP}} of Néel spirals for the Fe$_{19}$ 2a-[100] chains on Au/Nb(110). (a), the total DOS, including both electrons and holes, integrated along the chain plotted in the vicinity of the superconducting gap ($1.51$~meV), noted with green dashed lines. Calculated for Néel spirals with rotations angles changed in 5$^\circ$ steps between the ferromagnetic (0$^\circ$) and the antiferromagnetic (180$^\circ$) spin configurations, in the fully relativistic case, noted as $\text{SOC}=1$, representing the scaling factor for the SOC. (b) the electron LDOS at the Fermi energy along the 2a-[100] Fe$_{19}$ chain on Au/Nb(110) as a function of the Néel spiral rotation angle in the fully relativistic case ($\text{SOC}=1$). (d) and (e) are the same as \t(a) and (b), but with SOC scaled to 0. In panels (c) and (f) different cross sections are shown from (b) and (e) respectively, in order to better show the localization of the states, the lines are plotted with an offset of 200 arb. units.}
 \end{figure*}

Based on tight binding models\cite{Pientka2013}, it is to be expected, that spin spirals act along SOC to open up a minigap in the YSR band structure. To confirm this result, we studied the effect of different spin spiral states on the formation of the \red{ZEP} and the minigap in a series of calculations for Néel type spirals with spins rotating in the plane perpendicular to the surface ($\vec e_z$) and containing the chain ($\vec e_x$),
\red{described by the following local exchange-field as $\vec{B}(i)=|\vec{B}(i)| \left(\sin(\theta (i-9)) \vec e_x + \cos(\theta (i-9)) \vec e_z \right)$ on site $i$. This way}, the direction of the magnetic moment of the atom in the middle of the chain was fixed to point along the $z$ direction and the neighboring spins were rotated with respect to each other with an angle $\theta$ ranging from $0^\circ$ (FM) to 180$^\circ$ (AFM) in $5^\circ$ steps. 

The DOS (LDOS summed over all atoms in the chain) obtained for the different spirals as a function of the spiraling rotation angle in the fully relativistic case is shown in Fig.~\ref{fig_spiral}a. Probably the most interesting feature of this figure is the existence of a peak at zero energy which is present even in the FM state and remains undisturbed all the way until about $150^\circ$. Simultaneously we can see that there is no meaningful gap present around it in the FM case, as we noticed previously, however, as the spiraling angle is increasing, a minigap appears and increases in size from about 20$^\circ$. It keeps increasing until around $110^\circ$, where it reaches its maximum value of $0.25$ meV which is $16.5$ \% of the Nb gap. For larger spiraling angles the minigap starts to decrease and it collapses at around $150^\circ$ and then reopens again for even larger spiraling angles. While the minigap is open, from $20^\circ$ to $150^\circ$, there is a zero energy peak however when the gap closes and reopens at $150^\circ$ this peak disappears. Such behavior is usually a signal of a topological phase transition. It should be noted, that the zero energy peak is present even for the ferromagnetic chain, where the minigap is not yet fully opened. 

It was pointed out previously\cite{Marra2022review} that spin-orbit coupling plays an important role in the formation of MZMs. In our theory, it is possible to manipulate the Dirac equation in a way that the spin-orbit coupling term is scaled out, while all other relativistic effects, like the Darwin term and mass-velocity term is properly taken into account\cite{SOCscaleHubert}. In order to investigate the dependence of both the minigap and the \red{ZEP} on the SOC, we repeated our calculations with SOC scaled out. The results can be analyzed by comparing Fig.~\ref{fig_spiral}a and d. The calculations behind these figures are completely identical otherwise. Probably, the most prominent effect is, that without SOC the ferromagnetic state is gapped without a \red{ZEP} in it. When introducing a spiraling angle, this gap remains open until $40^\circ$ where it closes and reopens, however now a \red{ZEP} appears in it. The gap remains open with the \red{ZEP} until about $135^\circ$, where the gap closes and reopens again, this time without a \red{ZEP}. Consequently, even without SOC there is a large range of spirals where a \red{ZEP} can be observed. The spiraling angles where the minigap closes and reopens, also appear to be slightly different, when compared to the fully relativistic case. At the points where the gap closes (and reopens), a topological phase transition is expected, as we already discussed in the fully relativistic case, where only the second transition point is present. Our results without spin-orbit coupling is quite similar to what has been described in the context of previously studied simple models of MZMs previously\cite{NadjPerge2013, Martin2012, Pientka2013, Pientka2015}. 

\section{Spatial distribution of the zero energy peak}
\label{sec:localize}
It is known even from the original work of Kitaev\cite{kitaevchain}, that MZMs appear at the two ends of the chains. One should remember that the BCS pairing model leads to Cooper pairs which are formed by electrons with opposite momenta and spins thus mixing states from the region of the gap around the Fermi level. The coherence length is the extension of these wave packets in real space and proportional to $ 1/\Delta$. The frequently assumed physical picture is that the larger coherence length (smaller gap sizes) will be much more likely to cause larger localization length and thus hybridization of MZMs~\cite{Fleckenstein2018}. The results obtained here significantly changes this picture emphasizing the importance of spin-orbit effects and material-specific treatment. In order to examine the spatial extent of the \red{ZEP}, we plotted the value of the local DOS (LDOS) at zero energy along all spiral atoms for Fe/Au/Nb(110) in Fig.~\ref{fig_spiral}b and e, with and without SOC, respectively. Most convincingly we find that for spiraling angles where a \red{ZEP} is present, the states are localized to the atoms at the end of the chain, independently of SOC and the spiraling angle. One interesting case is the ferromagnetic Fe chain on Au/Nb(110) with SOC, where we can already see a \red{ZEP} sitting in a tiny but not perfect ``gap''. It is obvious from Fig.~\ref{fig_spiral}b, that there are zero energy states distributed along the entire chain and by the introduction of a spiraling angle, the states on the in-between atoms gradually disappear, and the states finally become localized to the ends of the chain around $20^\circ$. Therefore, even in the case of the ferromagnetic chain, there is a sharp state at zero energy, which continuously evolves into the end states of the gapped spirals exactly at the angle where the gap opens in Fig.~\ref{fig_spiral}a and d. In the ferromagnetic state, however, because the internal gap is closed, it is masked entirely by YSR states on the in-between atoms. To better examine the formation and localization of MZMs, we repeated the plot in Fig.~\ref{fig_spiral}c and f, where data from the figures above plotted for the $60^\circ$, $75^\circ$ and $110^\circ$ spirals separately. It can be seen that first of all the extent (the localization length) of the state changes with the spiraling angle, and is slightly different with and without SOC. Without SOC, the most delocalized state is obtained for the $110^\circ$ spiral, while with SOC this appears to be the most localized one. It can also be clearly observed, that with SOC, there is a small oscillatory tail to the side peaks, which overlap. Such behavior was seen in tight-binding models as well\cite{Rainis2013, Fleckenstein2018}, when the MZMs on the two ends overlapped. All in all, we can conclude that the MZMs extend roughly 4 atomic position, about 8 2D lattice constants or $26.4$ Angstroms. 

\section{The singlet and the triplet order parameter}
\label{sec:op}

\begin{figure*}
 \includegraphics[width=1\textwidth]{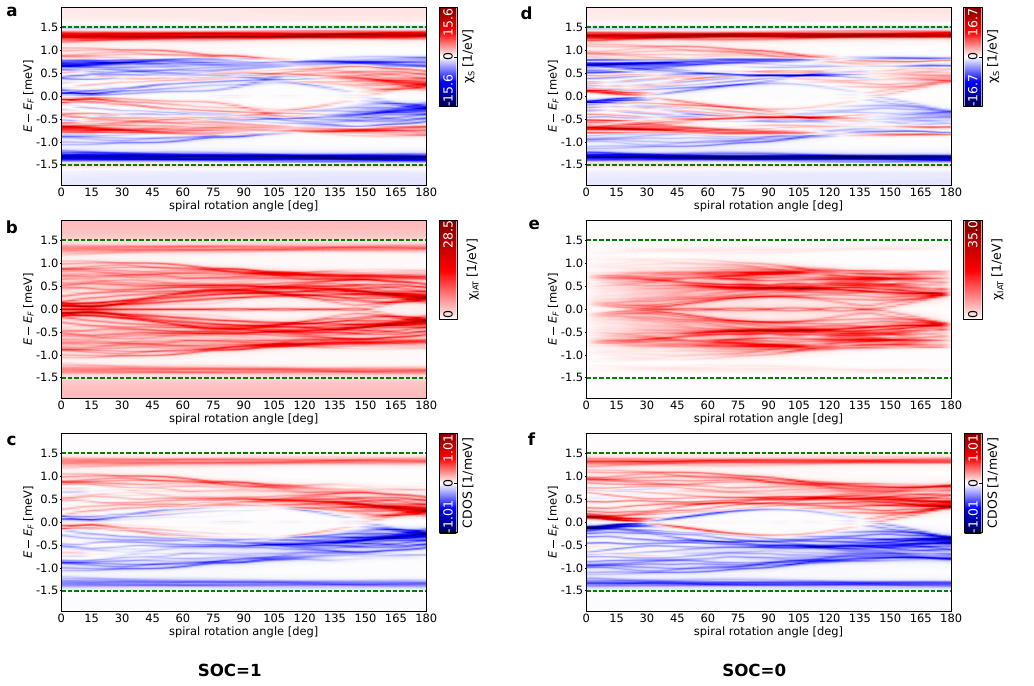}
    \caption{\label{fig_OP} The energy-resolved singlet(a), (d) and the norm of the IAT (b),(e) order parameters and the CDOS (c) and (f) of Néel spirals for the Fe$_{19}$ 2a-[100] chains on Au/Nb(110) integrated along the chain. The green dashed lines represent the superconducting gap of Nb ($1.51$~meV). Calculated for Néel spirals with rotations angles changed in 5$^\circ$ steps between the ferromagnetic (0$^\circ$) and the antiferromagnetic (180$^\circ$) spin configurations. The left column shows the fully relativistic case, noted as $\text{SOC}=1$, representing the scaling factor for the SOC, while the right column shows the scalar-relativistic case, $\text{SOC}=0$.}
\end{figure*}

In addition to the electron (and hole) densities, the KSDBdG equations provide us with a recipe to calculate the singlet and triplet OPs. 
In fact, the appearance of the superconducting state is manifested in the local Green's function matrix as finite elements in the electron-hole off-diagonal block. Hence, all the order parameters related to different pairing states shall be derived based on these elements. First, we define the following LDOS-like quantity to describe the energy resolution of the spin-singlet local OP\cite{Csire2016}:
\begin{equation}
    \chi_\text{S}(\varepsilon, n) = - \frac{1}{\pi} \Im~ \mathrm{Tr}_L~\mathcal{S}_s 
    \{ G^{nn,\text{eh}}_{Ls,L's'}(\varepsilon)\} \, ,
\end{equation}
where $\mathrm{Tr}_L$ denotes the trace in angular-momentum space, while $\mathcal{S}_s$ generates the spin-singlet,  $\mathcal{S}_s \{f(s,s^\prime)\}=\frac{\sqrt{2}}{2} \left(f(\frac{1}{2},-\frac{1}{2})-f(-\frac{1}{2},\frac{1}{2}) \right)$. The energy-resolved local singlet OP summed over the Fe atoms is shown in Fig.~\ref{fig_OP}a and d, with and without spin-orbit coupling, respectively. The singlet anomalous density shows very similar properties to the electronic DOS except from one characteristic difference, that there is no \red{ZEP}. This is a property of the Bogoliubov--de~Gennes theory, where the singlet OP is an odd function of the energy with respect to the energy zero level (this is a consequence of particle-hole symmetry) and therefore, it is zero at zero energy. Most non-zero energy states within the superconducting gap appear to have a non-zero OP, indicating a superconducting state. Some states however are such, that they are entirely electron-like or hole-like, which can be seen from the fact that the plot of the order parameter in Fig.~\ref{fig_OP}a and d does not exactly match the DOS plot of Fig.~\ref{fig_spiral}a and d, respectively. States which are entirely electron-like or hole-like are usually regarded as normal states, where the Cooper pairs are broken. It should also be mentioned that the magnitude of the singlet OP is quite small, which most likely comes from the rather uneven normal state density of states in spin channels which limits the formation of Cooper pairs and Andreev scattering. A larger contribution can be seen for the triplet OP, described below.

To further analyze the structure of the OP we consider the possibility of induced spin-triplet pairing since artificially constructed heterostructures were already proven to host spin-triplet Cooper pairs\cite{Robinson2010}. Here we aim for finding the dominant component of the induced triplet OP in real space, to scrutinize the behavior of the in-gap states. The fermionic nature of the electron implies that in the case of triplet pairing, the spatial component of the wave function has to be odd. In the context of a multi-band Hamiltonian for bulk systems, this allows the possibility of Even parity Odd orbital Triplet (EOT) states which has been shown to be responsible for the experimentally observed simultaneous appearance of magnetism and superconducting state in certain materials\cite{csiretriplet1,csiretriplet2}. In these cases, the translational invariance made it possible to introduce a proper parity operator for the whole system. However, since translational invariance is broken for surfaces and impurities, in order to avoid confusion with References~\onlinecite{ghosh2020recent,csiretriplet2, csiretriplet3} we shall adopt the term Internally Antisymmetric Triplet (IAT). It is expected that the relativistic Andreev scattering process (captured accurately by the generalized multiple scattering theory for the superconducting state) yields the largest contribution for IAT which is antisymmetric with respect to the orbital degrees of freedom. The common feature behind all these concepts is, that spin-orbit coupling do induce triplet pairing if a singlet pairing state already exists. This statement is easy to understand within our formalism, because during the solution of the relativistic Bogoliubov--de Gennes equation, a mixing occurs between the spin and orbital degrees of freedom together with the electron-hole character.
\red{This type of symmetry classification of Cooper pairs is important since we aim to distinguish these features from the odd-frequency spin triplet pairing which may also appear
in many artificial superconductor-magnet hybrid structures as presented in Ref.~\onlinecite{Cayao2020}.}
Therefore, we may also define a DOS-like quantity to account for the norm of the energy-resolved IAT order parameter (which is now a matrix in orbital indices):
\begin{equation}
    \overline{\chi}_{\text{IAT}} (\varepsilon) = 
     \sum_n \sum_{i=-1,0,1}
        - \frac{1}{\pi} \  \left| \left| \Im
              \mathcal{A}_L \mathcal{T}_s^i
              \{ G^{nn,\text{eh}}_{Ls,L's'}(\varepsilon)\}
            \right| \right|_\text{F},
\end{equation}
with the antisymmetrization in angular-momentum space, $\mathcal{A}_L \{ f(L,L^\prime)\}=$ $ \{ \frac{1}{2} \left( f(L,L^\prime)-f(L^\prime,L) \right) \}$, and the projections on spin-triplets, 
$\mathcal{T}_s^0 \{ f(s,s^\prime)\}=\frac{\sqrt{2}}{2} \left( f(\frac{1}{2},-\frac{1}{2})+f(-\frac{1}{2},\frac{1}{2}) \right)$, and 
$\mathcal{T}_s^{\pm 1} \{f(s,s^\prime)\}= f(\pm \frac{1}{2}, \pm \frac{1}{2})$, while $|| M ||_\text{F}$ denotes the Frobenius norm of matrix $M$.
This quantity accounts for the emergence of IAT pairing and
has been plotted in Fig.~\ref{fig_OP}b and e with and without SOC respectively.
It can be seen that for zero SOC and for a ferromagnetic (or anti-ferromagnetic)  configuration, the triplet OP is zero because the SOC is not inducing any mixing between the spin and the electron-hole indices. Additionally, there is a very small value of the singlet OP, and electron-hole mixing.

This is understandable and expected, based on the normal state DOS, and indicates that the electrons are almost entirely in the normal state.  In the case of SOC at its full value, even in the ferromagnetic chain there appears to be IAT (triplet) states present. By increasing the spiraling angle, the mixing between the spin channels become more substantial. The magnitude of the triplet OP is an order of magnitude larger than the singlet OP for all angles. In principle, there are all types of states present: zero pairing (normal state), a small amount of singlet and much more IAT pairing are simultaneously possible. Interestingly, a non-zero triplet OP can be seen at zero energy as well. This means that the \red{ZEP} is not only a state at zero energy, a state localized to the edge of the chain, but it is also an (IAT) triplet state. 

\section{Topological properties of the minigap} 
\label{sec:topology}

\begin{figure*}[htb]
\centering
\includegraphics[width=\linewidth]{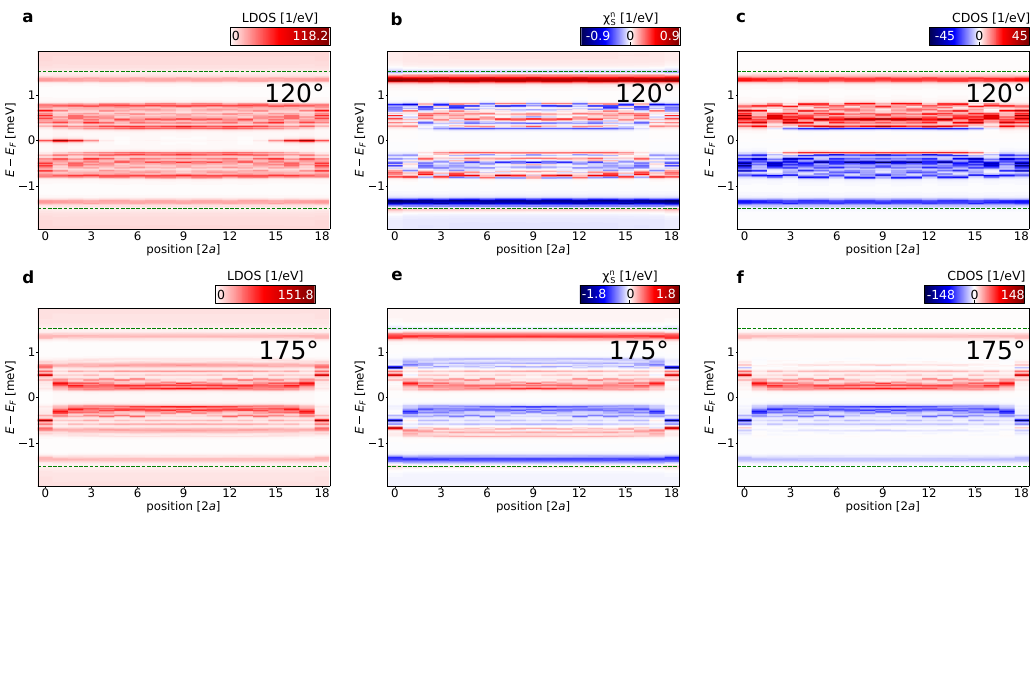}
    \vspace{-3cm}
\caption{\label{fig_topology} The real space structure of the LDOS and the energy-resolved singlet OP in the presence and in the absence of a \red{ZEP}. (a), the LDOS of the 2a-[100] Fe$_{19}$ chain on Au/Nb(110) in a Néel type spin spiral state with $120^\circ$ rotation angle. (b) the energy-resolved local singlet OP $\chi_\text{S}(\varepsilon,n)$ and (c) is the CDOS of the same spiral as in (a). (d), (e) and (f), the LDOS, the $\chi_\text{S}(\varepsilon,n)$ and the CDOS for the 175$^\circ$ spiral respectively. The superconducting gap of the Nb indicated by green dashed lines. The plots are include the values of the functions only for the Fe sites, the vacuum positions in between are neglected.}
\end{figure*}

Band inversion is a key signature of topological superconductivity, which can be observed if one investigates the band structure of infinite chains as a function of momenta. In finite chains it is not possible to observe band inversion in the strict definition of the term, as bands are not present. However, a real space solution and a momentum space solutions of the KSDBdG equations for infinite chains should carry the same information, and one may expect that the signatures of band inversion - if found in a \textbf{k}-space solution - also appear in the real space solution. Therefore it is expected that even a finite chain may carry the signatures of topological superconductivity - provided it is long enough - which then can be obtained from a real space calculation of the electronic states. 
In the superconducting state, the band inversion can be made visible by the construction of antisymmetric quantities with respect to the Fermi level. Based on the particle-hole symmetry, we found two such antisymmetric quantities which provide a clean visual picture of band inversion.  
One such quantity is the energy-resolved singlet OP, which is known to be antisymmetric\cite{Ketterson1999} with respect to the Fermi level. 
It can be seen in Fig.~\ref{fig_topology}, where the singlet OP is plotted, that while in the case of a $175^\circ$ spiral, where no \red{ZEP} is present (Fig.~\ref{fig_topology}d), the sign of the singlet OP is uniform on each atom, and the sign on either side of the minigap remains the same compared to the appropriate coherence peaks related to the bulk gap.
In stark contrast, for a $120^\circ$ spiral for example, where a \red{ZEP} is present at the edge atoms of the chain (Fig.~\ref{fig_topology}a),  it changes sign prematurely along the chain and the sign of the coherence peak of the minigap is the opposite of the appropriate bulk gap's coherence peak (Fig.~\ref{fig_topology}b) -- which is indicating band inversion and, consequently, a topological minigap. 
The signatures of band inversion and the appearance of zero energy edge states with the bulk-edge correspondence\cite{Liang2011} can be regarded as in silico evidence of the existence of MZMs and topological superconductivity.

A similar visualization of the band inversion can be obtained based on the quasi-particle charge density of states or in short the $\textrm{CDOS}$. In the superconducting state, assigning a unit charge $+e$ to the electron and $-e$ to the hole, the net quasi-particle charge density of states can be introduced as $\textrm{CDOS} (\varepsilon) = e \left( \textrm{DOS}_e (\varepsilon) - \textrm{DOS}_h (\varepsilon) \right) $ where $e$ is the electron charge.
Just as the density of states (DOS), CDOS can be local to a site or summed up for all the sites in the chain. 

Yet again, according to the particle-hole symmetry, 
the CDOS is antisymmetric with respect to the Fermi level ($\varepsilon = 0$), namely $ \textrm{CDOS} (\varepsilon) = -\textrm{CDOS} (-\varepsilon)$.
In Fig.~\ref{fig_OP}c and f one can observe the antisymmetric property of the CDOS and it can also be concluded that in spirals, where zero energy peaks are absent, the negative energy in-gap states have mainly electron-like character, while the positive energy in-gap states are mostly composed of hole-like states.
Fig.~\ref{fig_OP}c and f also illustrates that the CDOS, corresponding
to the induced minigap, changes sign prematurely at the minigap edge as the function of energy, which implies that the electron-hole character of these states are exchanged. 
As a function of the spiraling angle between $20^\circ$ to $150^\circ$, this feature happens simultaneously with the appearance of the zero energy state. Furthermore, the signatures of band inversion can be observed for lower spiraling angles as well, all the way to $0^\circ$, the ferromagnetic solution. This indicates that even though the gap is filled by in-between states as we described in section \ref{sec:localize}, the state at zero energy in the ferromagnetic state is an MZM, but it is more extended and overlapping.
A similar conclusion can be drawn from the singlet OP, shown in Figure \ref{fig_OP}a. 
One may also notice that neither the CDOS nor the OP shows the zero-energy states due to their antisymmetric property.
In fact, this process of transforming the internal "twisted" YSR states into the "untwisted" YSR states at the edges, gives rise to the emergence of zero energy edge states.
Altogether, these features are clear indications of band inversion and, consequently, a topological minigap for the spiraling angles where the zero energy states are observed at the ends of the Fe$_{19}$ chain.
Therefore, from this point on, they can be called Majorana Zero Modes due to the bulk-edge correspondence principle\cite{Liang2011}.

Another intriguing finding in Fig.~\ref{fig_OP}b-c (as the CDOS demonstrates it around 20$^\circ$ in the fully relativistic case) is that the gap closing and opening with the continuous change in spiraling angles may not necessarily imply a change in the topological behavior as well.
Such a change is usually considered as a rule of thumb in identifying topological superconductivity during experimental realizations. \red{On the other hand, there can be several reasons behind this observation including that in a complicated multiband system with SOC behaves differently from model predictions\cite{Ramires2022, csiretriplet2}, that the gap closing can cause the emergence of additional MZMs with an even number\cite{li2014topological,heimes2015interplay,dumitrescu2015majorana}, that the gap does not completely close but only tightens, or that this occurs for a finite (and insufficiently large) chain.}

\section{Conclusions}
To summarize, using first principles-based solution to the Dirac--Bogoliubov--de Gennes equations, we studied linear chains of Fe atoms placed on the surface of Au-covered Nb(110) in the superconducting state. We found that: (i) in agreement with experiments, a ferromagnetic state does not support a minigap around zero energy, (ii) in a spin spiral state however, a minigap emerges at about 20-degree spiraling angle with a spin-polarized zero energy state in it. This state bears all the signatures of being an MZM. It is localized to the two ends of the chain, it is a spin triplet state, and it is in a topological gap. At larger spiraling angles, the gap closes. (iii) We also showed that this state extends to about 4-5 atoms, and they more easily overlap in the presence of spin-orbit coupling. 
(iv) Moreover, our calculations revealed that the \red{MZMs bear the characteristic feature} of non-unitary internally antisymmetric triplet states, which are thought to be responsible for time-reversal symmetry breaking in a large class of bulk superconductors\cite{ghosh2020recent}.

In part 2 of our investigations, 
we will show that the Majorana Zero Mode we found in the present paper, is quite robust against various perturbations of the magnetic state. 
We will also explore potential routes to non-topological zero energy edge states, and combine different spin spirals to shed light on other fascinating - and potentially quite useful - phenomena: the shift of MZMs with changing spiraling angles
and topological fragmentation as a result of global phase shifts on the chain. 

\section*{Acknowledgements}
B. Ny., L.A., L.Sz. and B.U. acknowledge financial support by the National Research, Development, and Innovation Office (NRDI Office) of Hungary under Project Nos. FK124100, K131938 and K142652. B. Ny. and L.Sz. acknowledge support by the Ministry of Culture and Innovation  and the NRDI Office within the Quantum Information National Laboratory of Hungary (Grant No. 2022-2.1.1-NL-2022-00004). B. Ny. acknowledges the support by the ÚNKP-22-4-I New National Excellence Program of the Ministry of Culture and Innovation from the source of the NRDI Fund. The authors acknowledge KIFÜ for awarding us access to resources based in Hungary.

\section*{Appendix}
\subsection{The details of first-principles calculations}
The calculations were performed in terms of the Screened Korringa-Kohn-Rostoker method (SKKR), based on a fully relativistic Green's function formalism by solving the Dirac equation for the normal state\cite{Szunyogh1995} and the Kohn-Sham-Dirac-Bogoliubov-de Gennes (KSD-BdG) equation for the superconducting state within multiple scattering theory (MST)\cite{Csire2015, csire2018relativistic, Nyari2021}. The chains are included within an embedding scheme\cite{Lazarovits2002}, being an efficient method to address the electronic and magnetic properties or the in-gap spectra of real-space atomic structures without introducing a supercell. In calculations for the Fe/Au/Nb(110) system consist of seven atomic layers of Nb, a single atomic layer of Au and four atomic layers of vacuum between semi-infinite bulk Nb and semi-infinite vacuum. The Fe impurities are placed in the hollow position in the vacuum above the Au layer and relaxed towards the surface by 21\%, while the top Au layer is also relaxed inwards by 2\%. 
The relaxations are obtained from total-energy minimization in a VASP\cite{Kresse1996,Kresse1996a,Hafner2008} calculation for a single Fe adatom and are used in all of the calculations. For the potentials we employ the atomic sphere approximation (ASA), the normal state is calculated self-consistently in the local density approximation (LDA) as parametrized by Vosko \textit{et al.}\cite{Vosko1980} The partial waves within MST are treated with an angular momentum cutoff of $\ell_\mathrm{max}=2$. In the self-consistent normal state calculations, we used a Brillouin zone (BZ) integration with 253 $\mathbf{k}$ points in the irreducible wedge of the BZ and a semicircular energy contour on the upper complex plane with 16 points for energy integration. In order to take into account charge relaxation around the magnetic sites, the atomic chains are calculated with a neighborhood corresponding to 2 atomic shells or a spherical radius of $r=1.01~a$ around the Fe atoms. This way the atomic cluster used for embedding the Fe$_{19}$ chain contained 339 atomic sites altogether.
After having obtained the self-consistent potentials in the normal state, quantities in the superconducting state were calculated by a single-shot calculation by solving the KSDBdG equation with the experimental band gap $\Delta=1.51$~meV\cite{Beck2021} used as the pairing potential in the Nb layers\cite{Nyari2021}. The BZ integration for the host Green's function was performed by using an increasing number of $\mathbf{k}$ points with respect to the normal state, including 1891 points in the irreducible wedge of the BZ. A sufficient energy resolution of the LDOS in the superconducting gap is acquired by considering 301 energy points between $\pm 1.95$~meV with an imaginary part of 13.6~$\mu$eV related to the smearing of the resulting LDOS. 

\bibliography{main_bib}
\end{document}


\title{Supplementary Information for \\
Shiba band structure and topological edge states of artificial spin chains}


\author{Bendeg\'uz Ny\'ari$^{1,2}$} 
\author{Andr\'as L\'aszl\'o$^{3}$}
\author{G\'abor Csire$^{4,5}$}
\author{L\'aszl\'o Szunyogh$^{1,2}$}
\author{Bal\'azs \'Ujfalussy$^{3}$}

\maketitle

\noindent
$^1$Department of Theoretical Physics, Institute of Physics, Budapest University of Technology and Economics, M\H uegyetem rkp.~3., HU-1111 Budapest, Hungary

\noindent
$^2$ELKH-BME Condensed Matter Research Group, Budapest University of Technology and Economics, M\H uegyetem rkp.~3., HU-1111 Budapest, Hungary

\noindent
$^3$Wigner Research Centre for Physics, Institute for Solid State Physics and Optics, H-1525 Budapest, Hungary

\noindent
$^4$Materials Center Leoben Forschung GmbH, Roseggerstraße 12, 8700 Leoben, Austria.

\noindent
$^5$Catalan Institute of Nanoscience and Nanotechnology (ICN2), CSIC, BIST, Campus UAB, Bellaterra, Barcelona, 08193, Spain



\section{Details of the first-principles based treatment of artificial superconducting heterostructures}

\subsection{Self-consistent Kohn-Sham-Dirac-Bogoliubov-de Gennes equations}

The microscopic theory of inhomogeneous superconductors is based on the Bogoliubov-de~Gennes (BdG) equations~\cite{deGennes}. The relativistic generalization -- called Dirac--Bogoliubov--de~Gennes (DBdG) equations -- was established by the work of Capelle \emph{et al}.~\cite{Capelle1,Capelle2} that was later combined with density functional theory in Ref.~\onlinecite{csire2018relativistic}.

The relativistic order parameter -- with assuming a contact potential for the interaction -- is given by
\begin{equation}
 \chi(\vec{r})=\left< {\Psi}^T(\vec{r}) \boldsymbol{\boldsymbol{\eta}} {\Psi}(\vec{r})\right>,
\end{equation}
with the time-reversal matrix
\begin{equation}
 \boldsymbol{{\eta}}=
 \begin{pmatrix}
   0 & 1 & 0 & 0 \\
   -1 & 0 & 0 & 0 \\
   0 & 0 & 0 & 1\\
   0 & 0 & -1 & 0
 \end{pmatrix},
\end{equation}
and ${\Psi}(\vec{r})$ represents the four-component Dirac spinor field operator. The proper relativistic generalization leads to the following KSDBdG Hamiltonian written in Rydberg units ($\hbar=1$, $m=1/2$, $e^2=2$)
\begin{equation}
H_{\text{DBdG}}=
 \begin{pmatrix}
   H_D & D_{\text{eff}}(\vec{r}) \boldsymbol{\boldsymbol{\eta}} \\
   D_{\text{eff}}^\ast(\vec{r}) \boldsymbol{\boldsymbol{\eta}}^T & -H_D^\ast
 \end{pmatrix} \, ,
\end{equation}
where
\begin{equation}
 H_D=c \vec{\boldsymbol{\alpha}}\vec{p} + \left( \boldsymbol{\beta}-\mathbb{I}_4 \right) c^2/2+ \left( V_{\text{eff}}(\vec r)-E_F \right) \mathbb{I}_4 + \boldsymbol{\beta}  \vec{\boldsymbol{\Sigma}}\vec{B}_{\text{eff}}(\vec r) \, , 
\end{equation}
\begin{equation}
 \vec{\boldsymbol \alpha} =
 \begin{pmatrix}
  0 & \vec{\boldsymbol \sigma} \\
  \vec{\boldsymbol \sigma} & 0
  \end{pmatrix} \, , \qquad
 \boldsymbol \beta=
 \begin{pmatrix}
  \mathbb I_2 & 0 \\
  0 & -\mathbb I_2
  \end{pmatrix} \, , \qquad
 \vec{\boldsymbol \Sigma} =
 \begin{pmatrix}
   \vec{\boldsymbol \sigma} & 0 \\
    0&  \vec{\boldsymbol \sigma}
  \end{pmatrix} \, ,
\end{equation}
and $\vec{\boldsymbol \sigma} $ denotes the vector of the Pauli-matrices. By adopting the simple semi-phenomenological parametrization of the exchange-correlation functional described in Refs.~\onlinecite{Suvasini, Csire2015}, the effective electrostatic potential, exchange field and pairing potential can be written as
\begin{subequations}
\begin{eqnarray}
  V_{\text{eff}}(\vec r)   &=&  V_{\text{ext}}(\vec r) +
  \int \frac{\rho(\vec r')}{|\vec r - \vec r'|}  \dd^{~\!3}\!r' +
  \frac{\delta E^0_{xc}[\rho,\vec m]}{\delta \rho(\vec r)},\\
  \vec{B}_{\text{eff}}(\vec r)&=& \vec{B}_{\text{ext}}(\vec r) +
                              \frac{\delta E^0_{xc}[\rho,\vec m]}{\delta \vec m(\vec r)},\\
  D_{\text{eff}}(\vec r) &=& \Lambda \, \chi(\vec r),
\end{eqnarray}
\end{subequations}
where $\Lambda$ is the strength of the interaction responsible for superconductivity (which can be treated as an adjustable semi-phenomenological parameter), $V_{\text{ext}}(\vec r)$ is the external potential (e.g. the Coulomb attraction from the protons), $\vec{B}_{\text{ext}}(\vec r)$ is the external field, $\rho (\vec r)$ is the charge density, $\vec m (\vec r)$ is the magnetization density, $E^0_{xc}[\rho,\vec m]$ is the usual (local spin density approximation) exchange-correlation energy for normal electrons. The equations describe the relativistic generalization of BCS theory for inhomogeneous superconductors taking into account their realistic band structure involving magnetism and spin-orbit effects.

\subsection{Green's function method for solving the KSDBdG equations}

We use a Green's function based approach exploiting the real-space representation of the resolvent of the KSDBdG Hamiltonian, 
\begin{equation}
 \mathcal{G}(z)=\left( z \mathbb{I}
 -H_{\text{DBdG}}\right)^{-1}.
 \label{eq:res}
\end{equation}
The Multiple Scattering Theory (MST), i.e. the Korringa-Kohn-Rostoker (KKR) method\cite{Minr2018} gives direct access to the Green's function avoiding the step of calculating Kohn-Sham orbitals. In Ref.~\cite{csire2018relativistic} this technique was generalised for the solution of the KSDBdG equation with layered structure and in Ref.~\cite{Nyari2021} was further extended with embedding, allowing the treatment of magnetic impurities and nanostructures.

In MST, 
the potential is written as a sum of single-domain potentials centered around each lattice site, $n$, namely $V_{\text{eff}}(\vec r)= \sum_n V_n(r)$, $\vec B_{\text{eff}}(\vec r)= \sum_n \vec B_{n}(r)$, $D_{\text{eff}}(\vec r)= \sum_n D_{n}(r)$. The potentials treated within the atomic sphere approximation (ASA) are zero if $r=|\vec r_n|\geq S_n$, where $S_n$ is the radius of the Wigner-Seitz (WS) sphere that has the same volume as the atomic cell $n$.

In the relativistic case, we search the solutions of the KSDBdG equations in the following form
\begin{equation}
 \Psi(z, \vec r)= \sum_{Q}
 \begin{pmatrix}
  g^e_{Q}(z,r) \chi_{Q} (\hat r) \\
  \ii f^e_{Q}(z,r)\chi_{\overline Q} (\hat r) \\
  g^h_{Q}(z,r) \chi^*_{Q} (\hat r) \\
  - \ii f^h_{Q}(z,r)\chi^*_{\overline Q} (\hat r)
 \end{pmatrix},
\end{equation}
where $Q=(\kappa,\mu)$ and $\overline{Q}=(-\kappa,\mu)$ are the composite indices for the spin-orbit ($\kappa$) and magnetic ($\mu$) quantum numbers; $g^{e(h)}_{Q}(z,r)$ and $f^{e(h)}_{Q}(z,r)$ are the large and small components of the electron (hole) part of the solution, respectively. The spin-angular function is an eigenfunction of the spin-orbit operator $ K = \boldsymbol \sigma L +\mathbb I$,
\begin{equation}
 K \ket{\kappa \mu} = -\kappa \ket{\kappa \mu}.
\end{equation}
This basis set has the advantage that the corresponding matrix of the spin-orbit operator is diagonal and it also allows an artificial scaling of spin-orbit coupling for testing relativistic effects as described in Ref.~\cite{SOCscaleHubert}. For later purposes the following notations are also introduced: $\overline{l}=l-S_k$, and $S_k=\kappa/|\kappa|$ the sign of $\kappa$.

The magnetic field can be rotated to a local frame such that it points along the $\hat z$ direction. 
In the normal state it has two advantages: the least amount of coupling occurs between the states with different $\kappa,\mu$ quantum numbers, and there is no need to distinguish the left hand-side and right hand-side solutions.

With integration over the angular parts and using the orthonormality
of the Clebsch-Gordan coefficients
the radial KSDBdG equations for arbitrary magnetic field can be written as
\begin{equation}
\begin{split}
&
\begin{pmatrix}
z + E_F & c \left( \frac{\dd}{\dd r} + \frac{1}{r} - \frac{\kappa}{r} \right) & 0 & 0 \\
c \left( \frac{\dd}{\dd r} + \frac{1}{r} + \frac{\kappa}{r} \right) & z + E_F + c^2 & 0 & 0 \\
0 & 0 & E_F -z & c \left( \frac{\dd}{\dd r} + \frac{1}{r} - \frac{\kappa}{r} \right) \\
0 & 0 & c \left( \frac{\dd}{\dd r} + \frac{1}{r} + \frac{\kappa}{r} \right) & -z + E_F + c^2
\end{pmatrix}
\begin{pmatrix}
g^e_{Q}(z,r) \\
f^e_{Q}(z,r) \\
g^h_{Q}(z,r) \\
f^h_{Q}(z,r)
\end{pmatrix}
 = \\
&
\sum_{Q'}
\begin{pmatrix}
u^{++}_{Q Q'}(r) & 0 & \Delta_{Q Q'}(r)  & 0 \\
0 & u^{--}_{Q Q'}(r) & 0 & \Delta_{Q Q'}(r) \\
\Delta^*_{Q Q'}(r) & 0 & u^{++}_{Q Q'}(r)^* & 0 \\
0 & \Delta^*_{Q Q'}(r) & 0 & u^{--}_{Q Q'}(r)^*
\end{pmatrix}
\begin{pmatrix}
g^e_{Q'}(z,r) \\
f^e_{Q'}(z,r) \\
g^h_{Q'}(z,r) \\
f^h_{Q'}(z,r)
\end{pmatrix},
\end{split}
\end{equation}
where
\begin{equation}
u^{++}_{Q Q'}(r)= V(r) + \matrixelem{\chi_{Q}}{ \sigma_z B_z(r)}{\chi_{Q'}},
\end{equation}
\begin{equation}
u^{--}_{Q Q'}(r)= V(r) - \matrixelem{\chi_{\overline Q}}{ \sigma_z B_z(r)}{\chi_{\overline{Q}'}},
\end{equation}
\begin{equation}
 \Delta_{Q Q'}(r)=(-1)^{\mu'-\frac{1}{2}} ~S_{\kappa'} ~\delta_{\kappa \kappa'} \delta_{\mu~\! -\mu'} D(r).
\end{equation}
The last definition for the pairing potential matrix shows that the pairing interaction couples electrons with $\kappa,\mu$ quantum numbers to holes with $\kappa,-\mu$ quantum numbers. This is the direct consequence of our initial assumption that the pairing acts between Kramers pairs, namely between electrons and their time-reversed pairs (holes). The KSDBdG equations can be solved in a local frame with a predictor-corrector algorithm on logarithmic scale similarly, as it was done for the radial scalar relativistic BdG equations in Ref.~\onlinecite{Csire2015} and then it can be rotated into the direction of the exchange field.

The two most important quantities leading to the Green's function are the single-site $t$-matrix and the structure constants. Physically, the $t$-matrix describes scattering on the the single-site potential involving relativistic Andreev-scattering, while the structure constants contain all information about the crystal structure. This is one of the most amazing feature of the multiple scattering theory allowing the separation of scattering events and structural information. We shall denote the irregular and regular solutions of the KSDBdG equations inside the WS spheres by $\mathbf{J}_Q(z,\vec{r})$ and $\mathbf{Z}_Q(z,\vec{r})$ (matrices in electron-hole indices), respectively. The scattering solutions obey the following matching conditions (derived from the Lippmann-Schwinger equations as described in Refs. \cite{Csire2015, csire2018relativistic}.

\begin{subequations}
\begin{eqnarray}
    J_Q^{ab}(z,\vec{r}) &=& j_Q^a (z,\vec{r}) \delta_{ab},\\
    Z_Q^{ab}(z,\vec{r})&=& \sum_{Q'} j_{Q'}^{a} (z,\vec{r}) m^{ab}_{Q'Q}(z)
    -\ii p^a h_{Q}^{a}(z,\vec{r})\delta_{ab},\qquad
  \end{eqnarray}%
  \label{eq:GF_calcrel}%
\end{subequations}%
where $j_Q^a (z,\vec{r})$ and $h_{Q}^{a}(z,\vec{r})$ are spherical Bessel and Hankel type of solutions for the case of zero potential, respectively, and we introduced the inverse $t$-matrix, 
$m_{QQ'}^{ab}(z)=\left[\underline{t}^{-1}(z)\right]_{QQ'}^{ab}$.

As it was described in Ref.~\onlinecite{csire2018relativistic}, the relativistic real space structure constants should be derived from their non-relativistic counterpart $G_{0,LL'}^{\textrm{NR},ee,nm}(z) \delta_{ss'}$ with $L=(\ell,m)$ and $s$ being angular momentum and spin indices, respectively, by transforming it into the $\ket{\kappa \mu}$ basis, while the hole part of the relativistic structure constants can be constructed in the $\ket{\kappa \mu}^*$ basis using the non-relativistic formula\cite{Csire2015} $G_{0,LL'}^{\textrm{NR},hh,nm}(z)=-G_{0,LL'}^{\textrm{NR},ee,nm}(-z)$.

The following supermatrix formalism can be introduced for the scattering matrices, the matrices of the structure constant and the scattering path operator:
\begin{eqnarray}
 \mathbf{t}(z) &=& \{t^{n,ab}_{QQ'}(z) \delta_{nm} \},
 \label{eq:sh_t}\\
 \mathbf{G}_0(z) &=& \{G^{nm,ab}_{0,QQ'}(z) (1-\delta_{nm}) \delta_{ab} \},
 \label{eq:sh_G_0}\\
 \boldsymbol{\tau}(z) &=& \{\tau^{nm,ab}_{QQ'}(z) \},
 \label{eq:sh_tau}
\end{eqnarray}
where $\boldsymbol{\tau}(z)$ can be determined from the single-site t-matrix and the real space structure constant matrix 
\begin{equation}
 \boldsymbol{\tau}(z)=
 \left(
 \mathbf{t}(z)^{-1} -\mathbf{G}_0(z)
 \right)^{-1} \, .
  \label{eq:sh_tau_t}
\end{equation}

As in the normal state formalism, based on the expansions of the free-particle Green's function, 
the derivation of the one-particle Green's function  is straightforward and leads to the following formula,
\begin{equation}
\begin{split}
 \mathbf{G}(z ,\vec r,\vec {r'}) &=
   \sum_{QQ'} \mathbf{Z}_Q(z ,\vec r) \boldsymbol{\tau}_{QQ'}(z )
   \mathbf{Z}_{Q'}(z ,\vec{r'})^\times
 \\ &- \theta(r'- r) \sum_Q \mathbf{Z}_Q(z ,\vec r) \mathbf{J}_Q(z ,\vec{r'})^\times
 \\ &- \theta(r - r') \sum_Q \mathbf{J}_Q(z ,\vec r) \mathbf{Z}_Q(z ,\vec{r'})^\times,
\end{split}
  \label{eq:gf_formula}
\end{equation}
where $\theta(x)$ is the step function and
\begin{equation}
\begin{split}
\mathbf{Z}_Q(z ,\vec{r'})^\times
&= \mathbf{Z}_Q(z^*,\vec{r'})^\dagger \\
\mathbf{J}_Q(z ,\vec{r'})^\times
&= \mathbf{J}_Q(z^*,\vec{r'})^\dagger \, .
\end{split}
  \label{eq:Ztimes}
\end{equation}
stand for the left solutions of the KSDBdG equations.

\subsection{Treatment of layered systems and embedded clusters}
The formulas given above can be applied to surfaces and interfaces  quite straightforwardly following the idea of the so called Screened KKR (SKKR) formalism described in Refs.~\onlinecite{Szunyogh, Zeller}. In this formalism, it is made use of the 2D periodicity of the layers by introducing 2D lattice Fourier transformed version for the scattering path operator
\begin{equation}
 \boldsymbol{\tau}(z,\vec k_{||})=
 \left(
 \mathbf{t}(z)^{-1} -\mathbf{G}_0(z,\vec k_{||})
 \right)^{-1}.
  \label{eq:sh_tau_t2}
\end{equation}
To perform the inverse of the KKR matrix, a special reference system is used to obtain structure constants that are localized in real space. In the supermatrix formalism we used above, the screening transformation, described in details in Ref.~\cite{Zeller}, can be written in a way that is formally the same as it was presented in Sec.~III of Ref.~\cite{Szunyogh}. Thus the formalism can be derived for layered systems with two-dimensional periodicity and applied as the SKKR method prescribes.
It should be mentioned that a similar approach was developed for non-magnetic superconducting heterostructures with SOC effects in Ref.~\onlinecite{Rmann2022}.


The Embedded Cluster Method has been developed within multiple scattering theory to describe finite magnetic clusters of atoms in the superconducting state in Ref.~\cite{Nyari2021}, while a similar non-relativistic approach was presented in Ref.\cite{saunderson2022full} to treat  impurities embedded into a superconducting host.
In the knowledge of the layer resolved Green's function, the impurity problem can be solved by a Dyson equation. It can be shown that it is equivalent to modifying the SPO matrix to take into account the scattering due to the impurities. This is achieved by replacing the inverse scattering matrices $\mathbf {t}^{-1}_{\textrm{host}}(z)$ of the host atoms by those of the impurity or a cluster of impurities
$\mathbf {t}^{-1}_{\textrm{clus}}(z)$
to obtain the matrix of the scattering path operator at the impurity sites,
\begin{equation}
\boldsymbol {\tau}_{\textrm{clus}}(z) =
	\boldsymbol {\tau}_{\textrm{host}}(z) \left[
	\mathbb I - \left( \mathbf {t}^{-1}_{\textrm{host}}(z) -
        \mathbf {t}^{-1}_{\textrm{clus}}(z)
	\right) \boldsymbol {\tau}_{\textrm{host}}(z) \right]^{-1} 
\end{equation}
It should be noted that the embedding process requires
the calculation of the site off-diagonal elements of the scattering path operator and takes into account all relativistic Andreev scattering events both inside and outside the cluster.

By replacing $\boldsymbol {\tau}(z)$ with $\boldsymbol {\tau}_{\textrm{clus}}(z)$ (and the single-site scattering solutions with their impurity counterparts) in the Green's function formula Eq.~(\ref{eq:gf_formula}),
\begin{equation}
\begin{split}
 \mathbf{G}^{nm}(z ,\vec r,\vec {r'}) &=
   \sum_{QQ'} \mathbf{Z}^n_Q(z ,\vec r) \boldsymbol{\tau}^{nm}_{\textrm{clus},QQ'}(z )
   \mathbf{Z}^m_{Q'}(z ,\vec{r'})^\times
 \\ &- \theta(r'- r) \delta_{nm} \sum_Q \mathbf{Z}^n_Q(z ,\vec r) \mathbf{J}^n_Q(z ,\vec{r'})^\times
 \\ &- \theta(r - r') \delta_{nm} \sum_Q \mathbf{J}^n_Q(z ,\vec r) \mathbf{Z}^n_Q(z ,\vec{r'})^\times.
\end{split}
  \label{eq:gf_formula2}
\end{equation}

\subsection{The local Green's function matrix}

The density of states (DOS) of the system is defined as
\begin{equation}
    D(\varepsilon) = -\frac{1}{\pi} \int d^3r \Tr {\bf G}(\varepsilon+i0,\vec{r},\vec{r}) \, 
\end{equation}
where $\Tr$ denotes the trace in the electron-hole four-component, in total, eight-component vector space.
Restricting the integration to cell $n$, $V_n$, the DOS can be decomposed into cells and electron/hole components,
\begin{equation}
D(\varepsilon) = \sum_n 
\left( D^{n,e}(\varepsilon) + D^{n,h}(\varepsilon) \right) \, ,
\end{equation}
where
\begin{equation}
    D^{n,a}(\varepsilon) = -\frac{1}{\pi} \int_{V_n} d^3r \Tr {\bf G}^{nn,aa}(\varepsilon+i0,\vec{r},\vec{r}) \, 
\end{equation}
with $\Tr$ denoting the trace in four-component space only. We denote the above quantity as the local electron/hole DOS (LDOS).

In order to calculate order parameters describing the coupling between the electron and hole part of the Green's function, we need to resolve the LDOS by defining local Green's function (GF) matrix in electron-hole space
\begin{equation}
    G^{nn,ab}(\varepsilon) = -\frac{1}{\pi} \int d^3r \Tr {\bf G}^{nn,ab}(\varepsilon+i0,\vec{r},\vec{r}) \, .
\end{equation}
Inserting the real-space Green's function as provided by MST we can further resolve the local GF matrix according to $Q=(\kappa,\mu)$ indices,
By integrating the local radial Green's function within the atomic cell we arrive at our central quantity, the local Green's function matrix,
\begin{equation}
\begin{split}
        G^{nn,ab}_{QQ'}(\varepsilon) =
        \sum_{a'b'}\sum_{Q'',Q'''} F^{(ZZ)n,bb',aa'}_{QQ'',Q'Q'''}(\varepsilon)  \,
        \tau^{nn,a'b'}_{Q''Q'''}(\varepsilon)  - \sum_{a'}\sum_{Q''} 
        F^{(ZJ)n,ba',aa'}_{QQ'',Q'Q''}(\varepsilon)  \, ,
    \end{split}
\end{equation}
with the radial integral matrices,
\begin{equation}
    \begin{split}
F^{(ZZ)n,bb',aa'}_{QQ'',Q'Q'''}(\varepsilon) & =
\int_0^{S_n} r^2 dr \,
\left(g^{(Z)n,bb'}_{QQ''}(\varepsilon,r) \,
g^{(Z)n,aa'}_{Q'Q'''}(\varepsilon,r) +
f^{(Z)n,bb'}_{QQ''}(\varepsilon,r) \,
f^{(Z)n,aa'}_{Q'Q'''}(\varepsilon,r) \right) \, \\ 
F^{(ZJ)n,bb',aa'}_{QQ'',Q'Q'''}(\varepsilon) & =
\int_0^{S_n} r^2 dr \,
\left(g^{(Z)n,bb'}_{QQ''}(\varepsilon,r) \,
g^{(J)n,aa'}_{Q'Q'''}(\varepsilon,r) +
f^{(Z)n,bb'}_{QQ''}(\varepsilon,r) \,
f^{(J)n,aa'}_{Q'Q'''}(\varepsilon,r) \right) \, .
    \end{split}
\end{equation}
In the above definitions, $g^{(Z/J)n,ab}_{QQ'}(\varepsilon,r)$ and $f^{(Z/J)n,ab}_{QQ'}(\varepsilon,r)$ denote the radial parts of the large and small components of the regular/irregular scattering solutions, respectively.
After a transformation to the $(L,s) \equiv (\ell,m,s)$ basis in terms of Clebsch-Gordan coefficients  we obtain the matrix  $G^{nn,ab}_{LL',ss'}(\varepsilon)$. This quantity will be used to calculate the local densities of states and order parameters for embedded chains.

\section{Computational procedure}

The general algorithm for a complete calculation of an impurity cluster is then performed according to the following scheme. First, a series of normal state, conventional DFT calculations are performed within the same Green's function formalism, which is often referred to as Multiple Scattering Theory (MST). We begin with a bulk calculation to obtain self-consistent potentials and the Fermi energy for the bulk host (Nb). Then, a normal state surface calculation is done to obtain potentials and vacuum potential level for the semi-infinite host, in the current paper AuNb(110), where Au refers to a single atomic layer of Au on top of Nb(110). Still in the normal state, this is followed by an embedded cluster calculation to obtain the self-consistent potentials for the entire impurity cluster of 81 atoms. Here, the magnetic moments induced by the impurity cluster are relaxed as well. Once the normal state calculations are ready, and all the self-consistent potentials are obtained, the same steps are repeated in the superconducting state, however, without self-consistency. In these steps, we obtain the superconducting gap in the bulk, on the Au layer, and finally the LDOS, $\chi_\text{S}(\varepsilon, n), \chi_\text{IAT}(\varepsilon)$ quantities on the impurity cluster. It should be noted that we do not solve the problem of superconducting pairing interaction self-consistently, we set the effective pairing interaction $\Delta$ in the KSDBdG equations to a value that gives the experimentally observed gap for Nb in the bulk DOS calculation, and use the approximation of $\Delta=0$ for the Au, Fe and vacuum sites of the final system. We would like to mention, that because of the atomic sphere approximation used in our calculations, the somewhat artificial definition of the induced moments (magnetic moments that appear on non-magnetic sites due to interaction with magnetic sites), we tested our calculations by artificially setting them to zero. We found only a rather minimal effect of these induced moments, which is quite satisfactory. 

\newpage

\section{The normal state LDOS compared between the FM and $90^\circ$ spiral state}
\begin{figure}[H]
 	\centering
     \includegraphics[scale=1]{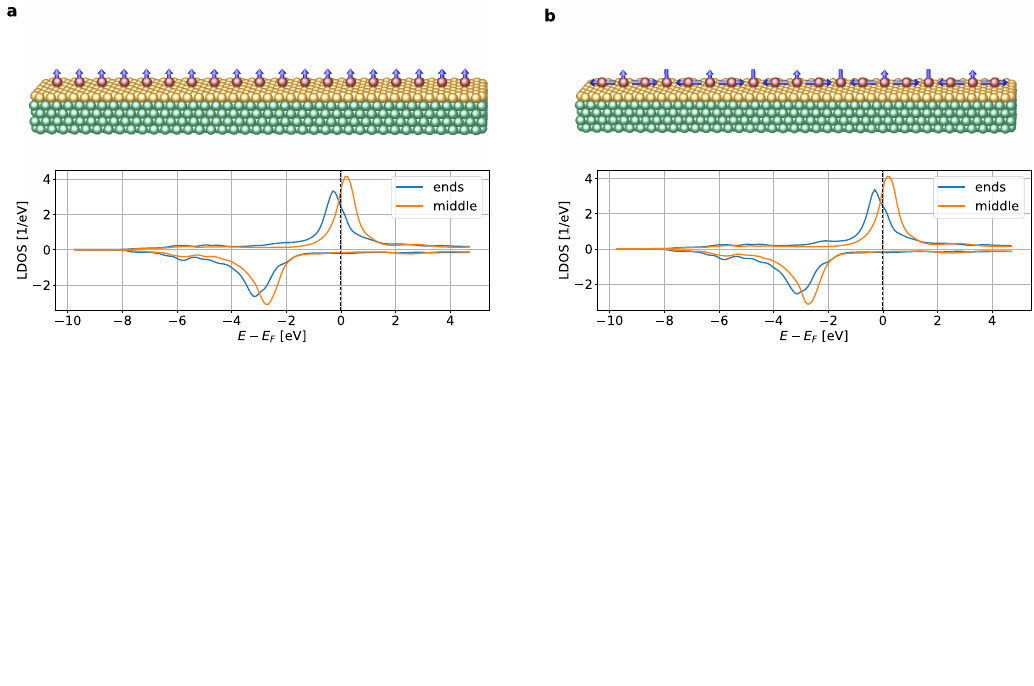}
     \caption{\label{sfig_normal}\textbf{The LDOS of the 2a-[100] Fe$_{19}$ chain on Au/Nb(110) in the normal state.}\\ \textbf{a}, the illustration of the 19 atomic 2a-[100] Fe chain on Nb(110) covered with a single monolayer Au in the ferromagnetic spin configuration and the spin resolved LDOS of the chain. \textbf{b} the illustration of the 90$^\circ$ spiral and the plot of the LDOS similarly as in \textbf{a}. The positive values are from the minority spin channel and the negative values are from the majority spin channel. The blue curves are calculated on the first atom of the chain and the orange are from the middle of the chain.}
 \end{figure}

In Fig.~1b of the main text we presented the spin resolved normal state LDOS of the ferromagnetic 2a-[100] $Fe_{19}$ chain on Au/Nb(110) for an atom at the end and at the middle of the chain. We claimed that the LDOS doesn't depend significantly on the spin configuration of the chain, however, neglected the plot for the 90$^\circ$ spiral. In Supp.Fig.~\ref{sfig_normal} we compare the LDOS in the two spin configuration and confirm that in contrast to the superconducting state in the normal state the two chains behave in a very similar way. The LDOS at the center atom is the same, the only visible difference is in the LDOS of the first atom at the Fermi energy where the curve is slightly flattened, but it is a very small difference doesn't imply the very different superconducting properties.

\newpage

\section{Superconducting DOS of the Au/Nb(110) layered system}
\begin{figure}[H]
 	\centering
     \includegraphics[width=\linewidth]{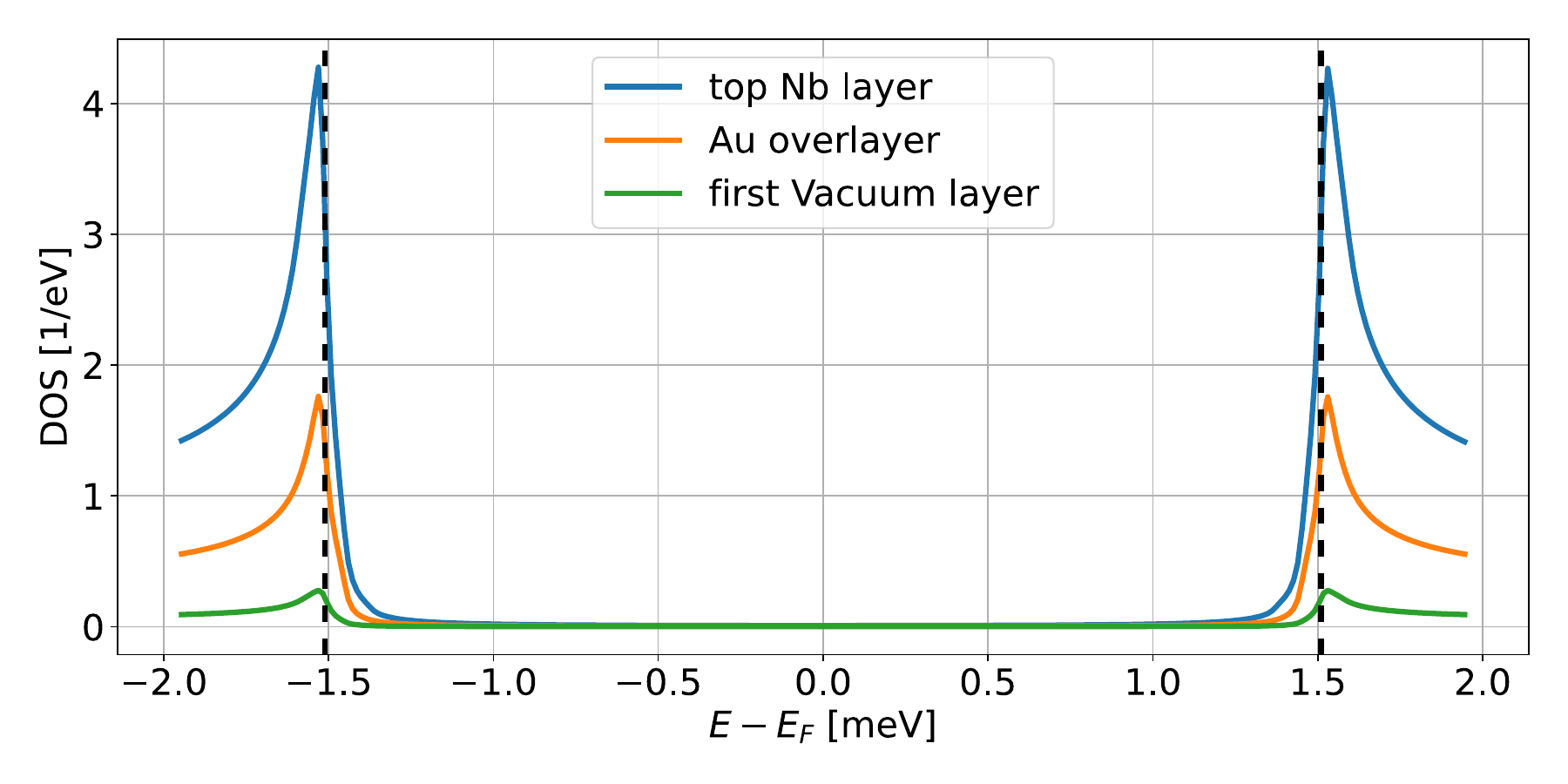}
     \caption{\label{sfig_layer}\textbf{The superconducting DOS of layered Au/Nb(110)}\\ The electron DOS in the vicinity of the superconducting gap of the Nb for the Au/Nb(110) system. }
 \end{figure}

The Au/Nb(110) substrate used as a host system for the calculations of 19 atomic Fe chains. One important property of this setup is that superconducting gap measured at the surface of the Au overlayer is indistinguishable from the superconduncting gap of the bulk Nb below it. In our calculation for this host system without the Fe impurities we applied the experimental value gap of the Nb $\Delta=1.51$ meV as a pairing potential on the Nb layers and on the semi-infinite bulk Nb below. For the Au layer and the vacuum we used $\Delta=0$. The results for the Nb-Au-vacuum interface shown in Supp.Fig.~\ref{sfig_layer}. The DOS is zero in the whole range of the gap and the coherence peak is at the same energy for each layer. It verifies that in the Au layer the induced superconducting gap due to the proximity to Nb is equals to the Nb gap providing a surface with similar superconducting properties to the pure Nb, but with enhanced SOC.  

\newpage
\section{The superconducting LDOS as a function of SOC}
\begin{figure}[H]
 	\centering
     \includegraphics[scale=1]{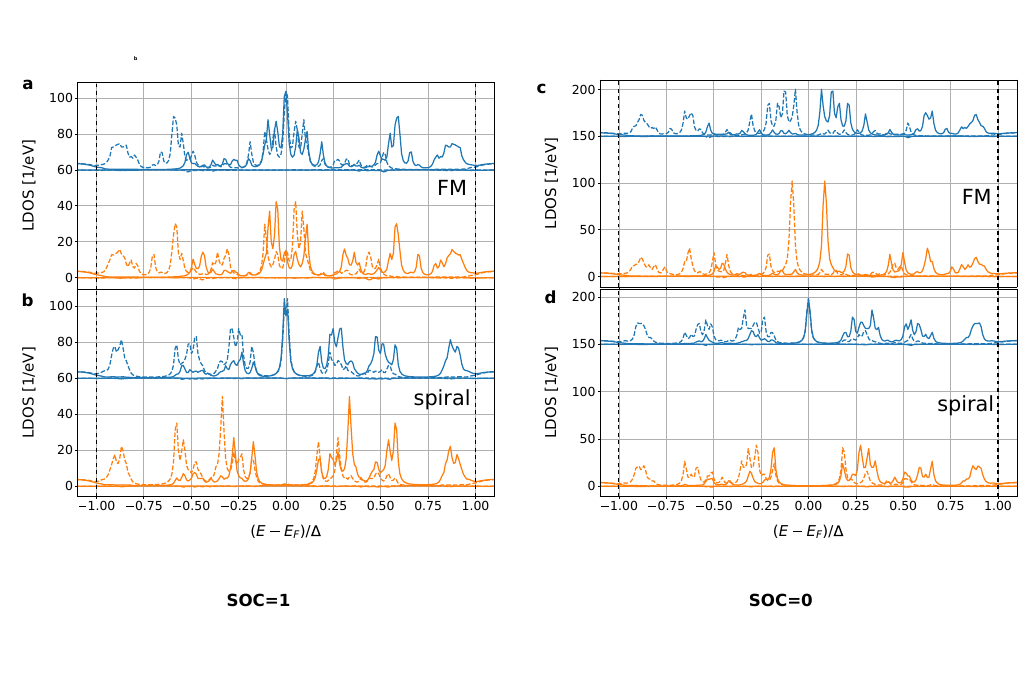}
     \caption{\label{sfig_scldos}\textbf{The LDOS of the 2a-[100] Fe$_{19}$ chain on Au/Nb(110) in the superconducting state as a function of SOC.}\\ \textbf{a}, the LDOS of the 19 atomic 2a-[100] Fe chain on Nb(110) covered with a single monolayer Au inside the superconducting gap in the ferromagnetic spin configuration with SOC. \textbf{b} same as \textbf{a} but for the 90$^\circ$ spiral. \textbf{c} and \textbf{d} is the plot of the same quantities as in \text{a} and \textbf{b} but with SOC scaled to 0. The positive values are from the minority spin channel while the negative values are from the majority channel. The solid lines are electron densities and the dashed line are hole densities, the blue curve is calculated on the fist atom and the orange is from the middle of the chains. The vertical dashed lines indicate the superconducting gap of the Nb (1.51 meV).}
 \end{figure}

We compare the LDOS of the FM and the 90$^\circ$ Néel spiral with and without spin orbit coupling as an extension to Fig.~1c and d from the main text. In each plots of Supp.Fig.~\ref{sfig_scldos} the main contribution to the LDOS is coming from the minority spin channel regardless of SOC. However an important effect of SOC can be observed by comparing the FM cases. In the SOC=0 case there is a negligible electron-hole mixing, however if we are taking into account SOC a significant electron-hole mixing appears in the vicinity of the Fermi energy, which is still present in the spiral cases. This property of the LDOS also a sign of band inversion, produced by SOC or a spiral state, and an indicative sign of topological superconductivity.

\textbf{Supplementary References}
\bibliography{supp_bib}